\documentclass[12pt]{article}
\usepackage{graphicx}
\usepackage{amsopn}
\usepackage{amsfonts}
\usepackage{amsmath}
\usepackage{latexsym}

\title{Intermittency effects in Burgers equation
driven by thermal noise.}

\author{I.V.Kolokolov\\
Budker Institute of Nuclear Physics,\\
630090, Novosibirsk.\\
kolokolov@inp.nsk.su}
\date{}

\begin{document}

\maketitle
\begin{abstract}
For the   Burgers equation
driven by thermal noise
leading asymptotics of pair and high-order correlators
of the velocity field are
found for finite times and large distances.
It is shown that the intermittency takes place: some correlators
are much larger than their reducible parts.
(Pis'ma v ZhETF, v.71, n.1, January 2000, in press, English Translation in
JETP Letters, v.71, January 2000).
\end{abstract}

Intermittency implies strong non-Gaussianity of statistics of
fluctuating fields. This phenomenon is shown by hydrodynamical systems
in a state of developed turbulence\cite{Frish,CFKL,GK}.
In such far from equilibrium
situations intermittency appears as prevalence of irreducible parts of
some fourth order simultaneous correlators over reducible ones.

As for thermal equilibrium, irreducible parts of simultaneous correlators
of local fluctuating fields turn out to be of the same order as their
Gaussian parts even in critical region. This property is inherent in
the renormalization group method which takes care of interaction of
fluctuations through renormalization of local field and effective
Hamiltonian  \cite{PatP}.

In the recent paper \cite{Le} V.V.Lebedev disclosed
that the picture can change
drastically when we pass to time-dependent correlations of thermally
fluctuating quantities. He found that in the low-temperature
phase of two-dimensional systems of the Berezinskii-Kosterlitz-Thouless
kind different-time correlation functions of the vortex charge density
may greately exeed their own Gaussian part.  In the same paper \cite{Le}
the physical cause of such the behaviour is pointed out: at low
temperatures the non-simultaneous correlation function of every order
in vicinity of a given poin are defined by a single rare fluctuation.
One can conclude from this interpretation that the intermittency effects
may emerge in equilibrium dynamics of a wide range of systems.

In the present paper I consider one-dimensional velocity field evolving
according to the Burgers equation with the thermal noise term:
 \begin{equation}
\label{ur}
u_t+uu_x-\nu u_{xx}=\xi(t,x).
\end{equation}
Here $\nu$ is the dissipation constant and
$\xi(t,x)$ is random noise with Gaussian statistics and the pair
correlator:
 \begin{equation}
\label{shum}
\langle\xi(t,x)\xi(t_1,x_1)\rangle=-\nu\beta^{-1}
\delta^{\prime\prime}(x-x_1)
\delta(t-t_1).
\end{equation}
We will consider $\nu$ as being vanishingly small.
The parameter $\beta$ plays the role of inverse temperature, so
the simultaneous stationary distribution function
${\cal P}[u]$ has the Gibbs form:
 \begin{equation}
\label{ravn}
{\cal P}[u]={\cal N}\exp\left\{-\beta{\cal F}[u]\right\},
\quad {\cal F}[u]=\int \,dx\,u^2(x).
\end{equation}
Here ${\cal N}$ is the normalization constant.
The expression:
 \begin{equation}
\label{odpar}
\langle u(t,x)u(t,x')\rangle=(2\beta)^{-1}\delta(x-x').
\end{equation}
following from (\ref{ravn}) correspond to a total absence of velocity
correlation in spatially distant points in a given time moment.
In the present paper some asympotics of various
non-simultaneous correlators of the field
$u(t,x)$ are found. The results obtained here show presence of intermittency
effects in the equilibrium dynamics of the system (\ref{ur}).

Dynamical scaling exponent $z=3/2$
for the problem (\ref{ur})-(\ref{shum}) was discovered
in the paper \cite{FNS} using dimensional analysis and utilizing the
Galilean invariance.
In the paper \cite{LeL} the absence of logarithmic divergencies
for the spectrum
$\omega\propto k^{3/2}$ was checked in every order of renormalized
perturbation theory. Thus the function
$F_2(T,x)=<u(T,x)u(0,0)>$ has   $\beta x^3/T^2$
as a dimensionless argument. First we find the main (exponential)
part of the asymptotics of the function
$F_2$ at $\beta x^3/T^2\gg 1$
and $\nu\to 0$.  The latter limit means that the diffusion
cannot set up correlation of velocity in the points
$0$ and $x$ in a time $T$. The role of the noise in dynamics
on the time interval
$(0,T)$ is also negligible. Thus we can consider
$u(0,y)$ as a functional of $u(T,x)$ and vice versa. The velocity
statistics at the time moment $T$  is Gaussian what allows us to represent
the non-simultaleous correlator $F_2$ in the form:
 \begin{equation}
\label{parvar}
F_2(T,x)=(2\beta)^{-1}
\left\langle\frac{\delta u(0,0)}{\delta u(T,x)}\right\rangle.
\end{equation}
The variational derivative $\Theta(t,y)= \delta u(t,y)/\delta u(T,x)$
for $\nu\to 0$ satisfies the continuity equation:
 \begin{equation}
\label{urnep}
\Theta_t+u\Theta_y+u_y\Theta=0,
\end{equation}
and the condition $\Theta(T,y)=\delta(x-y)$.
This Cauchu problem is solved by the characteristic method and
we arrive to the expression for
$F_2(T,x)$:
 \begin{equation}
\label{parvare}
F_2(T,x)=(2\beta)^{-1}\langle\Theta(0,0)\rangle=
(2\beta)^{-1}\left\langle\delta\left(x-y(T)\right)
\left(\frac{\partial y(T,\zeta)}{\partial\zeta}\right)_{\zeta=0}
\right\rangle,
\end{equation}
(see \cite{ZeM}).  Here $y(T,\zeta)$ is coordinate of the Lagrange particle
started at the instant
$t=0$ from the point $\zeta$:
 \begin{equation}
\label{lagr}
\dot{y}=u(t,y),\quad y(0,\zeta)=\zeta,
\end{equation}
and $y(T)=y(T,0)$. If $u(t,y)$ is discontinuous, then the equation
(\ref{lagr}) requires a regularization. We use physically evident
condition that the particle on a shock wave front moves with the
velocity of this front. Its formal justification starting from a
finite small viscosity can be found in \cite{Bern}.

The expression (\ref{parvare}) tells us that the correlator
$F_2$ in the limit being considered is determined by a most probable
initial velocity fluctuation
$u_0(y)$ which, evolving, carries out the particle
from the point $0$ to the point  $x$ in the time $T$.
The probabilities of initial configurations are defined by
the functional (\ref{ravn}). The desired optimal fluctuation
$u_0(y)$ minimizes ${\cal F}[u_0]$ with the condition
$y(T)=x$. Let us show that it is the linear profile:
 \begin{equation}
\label{prol}
u_0(y)=u^*_0\equiv x/T-y/T,\;0<y<x,\quad u_0(y)=0,\;y<0,\,y>x.
\end{equation}
First, it is evident that the function
 $u_0(y)$  must have a maximum at $y=0$. It is also easy to understand
the equality $u_0(y)$  to 0 for $y<0$ and $y>x$. Indeed, difference
$u_0(y)$ from zero outside the interval $(0,x)$ does not affect
the trajectory $y(t)$ but  ${\cal F}[u_0]$ increases.
The left edge of the distribution
$u(t,x)$ for $\nu\to 0$ will be straight line with the slope
$\sigma=1/t$. Such the time dependence can be checked by direct
substitution into Burgers equation; see also  \cite{Gurb}.
At  $t=T$ the coordinate of the most rapid particle
will be equal to $x$. Coordinates of the other particles
from the interval $(0,x)$ will be precisely the same.
Thus, for the class of initial distributions $u_0(y)$
described above the plot of the final function
$u(T,y)$ has a form of triangle:
 \begin{equation}
\label{prot}
u(T,y)=y/T,\;0<y<x,\quad u_0(y)=0,\;y<0,\,y>x.
\end{equation}
Now let us note that from the Burgers equation it follows:
 \begin{equation}
\label{ner}
d{\cal F}[u(t,y)]/dt=-2\nu\int\,dy\,u_y^2\leq 0,
\end{equation}
what means that:
 \begin{equation}
\label{n}
{\cal F}[u_0(y)]\geq {\cal F}[u(T,y)].
\end{equation}
This inequality becomes strict one even for  $\nu\to 0$ if
shock waves were formed during the evolution. Consequently,
the minimal admissible value of the functional ${\cal F}$
is equal to:
 \begin{equation}
\label{min}
{\cal F}[u(T,y)]=x^3/3T^2.
\end{equation}
The value of ${\cal F}$ on the function $u^*_0(y)$ coincides
with (\ref{min}). The exlusion of shocks in the time interval $(0,T)$
justified above makes the expression (\ref{prol}) the only possible.

Probability of the initial fluctuation (\ref{prol}) proportional
to $\exp\left(-\beta{\cal F}[u_0(y)] \right)$ defines the exponential
part of the asymptotics of the pair correlator $F_2$:
 \begin{equation}
\label{prk}
F_2(T,x)\sim\exp\left(-\frac{\beta x^3}{3T^2}\right).
\end{equation}
It is worth noting that the multipiler
$(\partial y(T,\zeta)/\partial \zeta)_{\zeta=0}$ of the
$\delta$-function in the formula
(\ref{parvare}) vanishes on the configuration (\ref{prol}),
but it becomes different from zero under small variation of $u_0(y)$.
In other words, this factor, along with unknown pre-exponential
factor in the expression
(\ref{prk}) as a whole, is determined by integration
over variations $\delta u$ of the initial velocity field with respect
to $u^*_0(y)$. The essential values of $\delta u$ are small comparing
with $u^*_0(y)$; the parameter of this smallness is $(\beta x^3/T^2)^{-1}$.
However, the integration over
$\delta u$ cannot be reduced to the Gaussian one even
in the limit $\beta x^3/T^2\gg 1$. The point is that at $\nu\to 0$
the functional ${\cal F}[u]$ is not analytical  on the class of
initial velocity fields $u(y)$ obeying the constraint $y(T)=x$.
The variation $\delta {\cal F}$ turns out to be of the first order
in $\delta u$ despite that inequality  $\delta {\cal F}\geq 0$
holds. ${\cal F}[u]$ can be expanded in functional Taylor series in
$\delta u$ for $\delta u\ll \nu/x$ only. The corresponding analysis
will be given in another paper and here we restrict ourselves
to exponential asymptotics.

Noting that the initial linear profile (\ref{prol}) tranfers all the
internal point of the interval $(0,x)$  by the time $t=T$
into the point $x$ we conclude that up to pre-exponential factor:
 \begin{equation}
\label{mngk}
F_{n+2}=\left\langle u(T,x)\prod\limits_{j=1}^n u(0,y_j)\;u(0,0)\right\rangle
\sim F_2(T,x)\sim\exp\left(-\frac{\beta x^3}{3T^2}\right).
\end{equation}
Here $0<y_1<y_2\dots<y_n$. It is obvious that the reducible part of this
correlator is equal to zero. The same initial fluctuation
$u^*_0(y)$ determines leading asymptotics of the correlator
$\Phi_4=\langle u(T,x)u(T,x+a_1)u(0,a)u(0,0)\rangle$ at
$0<a<x$ and $0<a_1\ll a$:
 \begin{equation}
\label{phich}
\Phi_4\sim\exp\left(-\frac{\beta x^3}{3T^2}\right)\gg \Phi_{4,Gauss}
\sim \exp\left(-\frac{2\beta x^3}{3T^2}\right).
\end{equation}
Here $\Phi_{4,Gauss}$ designates reducible part of $\Phi_4$.
To find $\Phi_4$ as a function of the parameter $a$ it is
necessary to analyse the evolution of the perturbed linear profile.
In this case the shock waves formation becomes inevitable what makes
the problem more difficult. It is worth adding that
$a$-dependence  of the correlator $\Phi_4$ may be related
directly to the probability distribution function of velocity
field gradients. In \cite{Sin,BFKL} it was shown that the latter
id defined by forming shocks. Proportionality of asymptotics
of high order correlation functions to the asymptotics of pair
correlator is characteristic for turbulent-like problems
and in such the context was noted in \cite{LvF}.

Correlation functions of the field
$u(t,x)$ may be represented in a form of functional integrals
(see e.g., \cite{KacL}). In the present paper such the integrals
were computed in essence by the saddle-point method with the
parameter $\beta x^3/T^2\gg 1$ contained into the object to be
averaged, but not inherent to the action.
This approach goes back to the works of I.M.Lifshits (see in
\cite{Lif}). Later it was generalized to find high order correlators
in equilibrium \cite{Li} and strongly non-equilibrium problems
\cite{FE,FaKLM,GM,BFKL,Che,BF,BL}. Optimal fluctuation is called
also as an instanton by analogy with quantum field theory.
In the paper \cite{KM} long-time asymptotics of the current
autocorrelation function has been computed for a disordered contact
and the large observation time was used as a saddle-point parameter.

I am greately indebted to Michael Chertkov and Vladimir Lebedev
for numerous discussions and advices. I am grateful to Gregory
Falkovich and Thomas Spencer for backing investigations leading to this
work. I would like to appreciate Michael Stepanov for useful remarks.

\end{document}